\documentclass{article}

\usepackage[preprint]{neurips_2024}

\usepackage{graphicx}

\usepackage[utf8]{inputenc} % allow utf-8 input

\usepackage[T1]{fontenc}    % use 8-bit T1 fonts
\usepackage{hyperref}       % hyperlinks
\usepackage{url}            % simple URL typesetting
\usepackage{booktabs}       % professional-quality tables
\usepackage{amsfonts}       % blackboard math symbols
\usepackage{nicefrac}       % compact symbols for 1/2, etc.
\usepackage{microtype}      % microtypography
\usepackage{xcolor}         % colors

\title{Neural Signal Operated Intelligent Robot: Human-guided Robot Maze Navigation through SSVEP}

\author{%
  Jiarui Tang\\
  Qingdao Academy\\
  Qingdao, Shandong, China 266111\\
  \texttt{often0011@outlook.com} \\
   \And
  Tingrui Sun \\
  Qingdao Academy \\
  Qingdao, Shandong, China 266111 \\
  \texttt{} \\
  \AND
  Siwen Wang \\
  Qingdao Academy \\
  Qingdao, Shandong, China 266111 \\
  \texttt{wangsiwen@qdzx.net} \\
}

\begin{document}

\maketitle

\begin{abstract}
Brain-computer Interface (BCI) applications based on steady- state visual evoked potentials (SSVEP) have the advantages of being fast, accurate and mobile \citep{norcia2015steady}. SSVEP is the EEG response evoked by visual stimuli that are presented at a specific frequency, which results in an increase in the EEG at that same frequency. 
In this paper, we proposed a novel human-guided maze solving robot navigation system based on SSVEP. By integrating human's intelligence which sees the entirety of the maze, maze solving time could be significantly reduced. Our methods involve training an offline SSVEP classification model, implementing the robot self-navigation algorithm, and finally deploy the model online for real-time robot operation. Our results demonstrated such system to be feasible, and it has the potential to impact the life of many elderly people by helping them carrying out simple daily tasks at home with just the look of their eyes.
%Robot maze navigation has always been a challenge in computer algorithms. To develop a robot that can make real-time decisions and handle unexpected events posts a great challenge.     % The abstract paragraph should be indented \nicefrac{1}{2}~inch (3~picas) on
  % both the left- and right-hand margins. Use 10~point type, with a vertical
  % spacing (leading) of 11~points.  The word \textbf{Abstract} must be centered,
  % bold, and in point size 12. Two line spaces precede the abstract. The abstract
  % must be limited to one paragraph.
\end{abstract}

%\section{Submission of papers to NeurIPS 2024}
\section{Introduction}
Brain-computer Interface (BCI) applications based on steady-
state visual evoked potentials (SSVEP) have the advantages of being fast, accurate and mobile. The SSVEP is the EEG response evoked by visual stimuli that are presented at a specific frequency, which results in an increase in the EEG at that same frequency \citep{yang2022steady}. The majority of SSVEP applications focus on the spelling task. Usually the subjects are asked to stare at a group of letters flashing at different rate on the screen, and EEG data is simultaneously recorded to decode which letter the subject is currently looking at. Many researchers in the field are constantly working on expanding the number of classes for classification as well as the accuracy of the classification algorithm \citep{zhao2021filter} \citep{chen2023transformer}.
%As a group of high school students, we know that it would be extremely challenging to come up with a better algorithm for SSVEP classification. Thus, we decided to slightly modifying an existing algorithm and focus on expanding the application side of SSVEP. 
Robot maze navigation has always been a challenging field in computer algorithms. Being able to design a robot that is capable of making real-time decisions and solve unexpected problems post a great challenge \citep{magallan2022implementation}. The current directions of robot maze navigation has been focused on equipping the robot with more sensors to acquire more information and improving path finding algorithms \citep{aqel2017intelligent}. Most of the existing algorithms rely solely on the robots to make its own decision, such method could be advantageous in applications which human supervision cannot be provided. However, in some applications in which a human supervisor is available and time is a key factor, it may be advantageous for human and robot collaboration \citep{vysocky2016human}. In this paper, we propose a novel robot maze navigation system with contactless human guidance through SSVEP. 
%The robot car is equipped with basic path finding algorithms, and when a decision has to be made, usually at a cross-section, the robot stops and asks for human guidance. The operator could give instructions to the robot car by staring at one of the threes flashing LEDs equipped at the front, left and right side attached on the robot, each LED corresponds to the command "move forward", "turn 90 degrees left" and "turn 90 degrees right". 
Combining with the human's intelligence that sees the entirety of the maze, the robot could navigate its way out of the maze more efficiently. We believe such a human machine collaboration scheme could be implemented at elderly homes to help elderly people carry out simple daily tasks such as object retrieval, taking out trash and assist with other activities.

% Inspired by ...anime/movie, we decided to built a robot car that can be controlled with the mind, using SSVEP qw

%Please read the instructions below carefully and follow them faithfully.

\section{Method}

\subsection{Overview}
\begin{figure}
    \centering
    \includegraphics[width=0.75\linewidth]{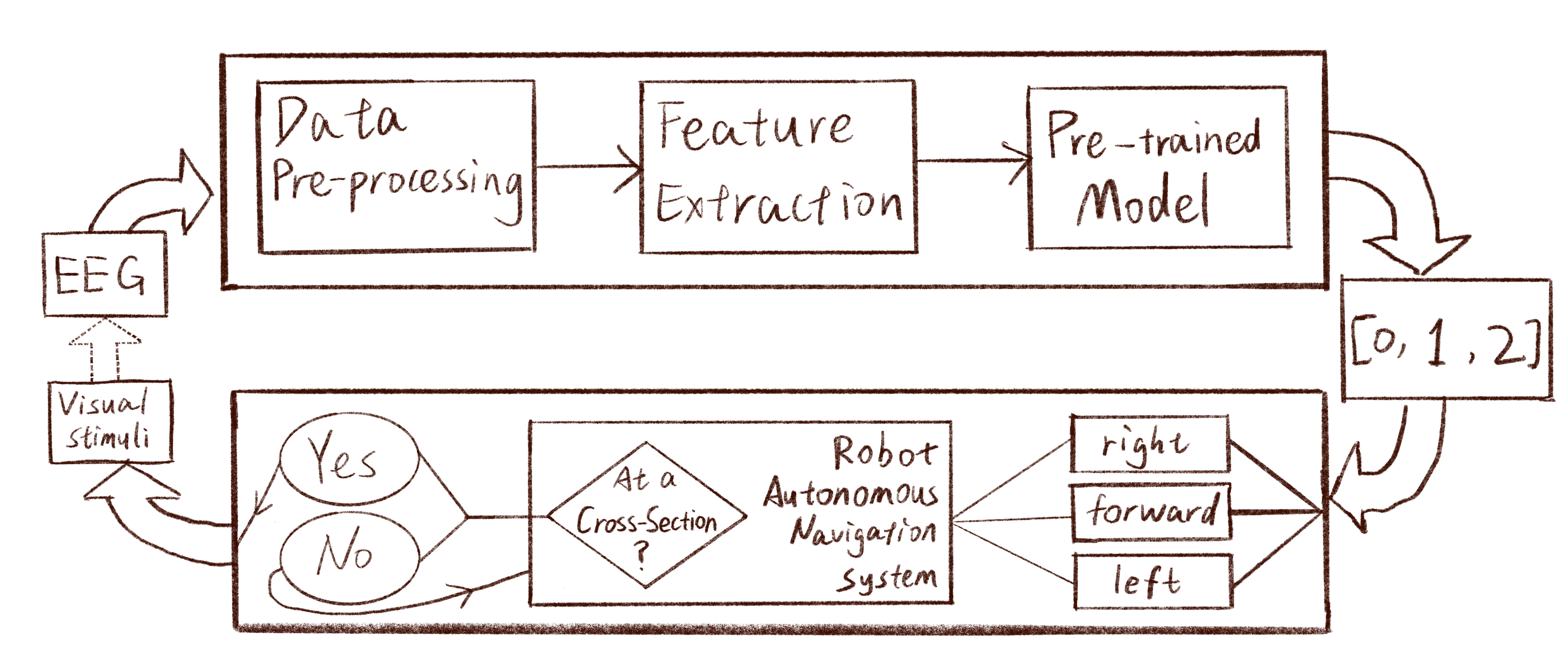}
    \caption{Closed-loop block diagram of the neural operated robot system}
    \label{fig:blockdia}
\end{figure}

% Talk about our task in general, show the closed-loop block diagram here. 
Our task to build a human- guided intelligent maze solving robot involves two major part: first, it requires us to train a classification model to correctly identify the SSVEP signal the human operator gives to the robot; 
%In the current prototype version, we implement only three commands: "move forward", "turn 90 degrees left" and "turn 90 degrees right" represented by 3 LEDs flashing at 9.25 Hz, 11.25 Hz and 13.25 Hz; 
the second part involves implementing an autonomous path finding algorithm, so that the robot only asks for human instruction when needed. 
%In the current version, the robot only asks for human intervention when it can turn left or right. 
Figure \ref{fig:blockdia} show a graphical representation of how both system works together.

\subsection{Offline training}

\subsubsection{Dataset}
Our training data is from the materials published online by Masaki Nakanishi (\url{https://github.com/mnakanishi/12JFPM_SSVEP/tree/master/data}). 

%The dataset contains 8-channel EEG data from 10 subjects conducting a SSVEP gazing task. The experiment consists of 15 blocks and there are 12 trials in each block. Each subject was asked to gaze at a visual stimulus marked randomly by the computer among 12 stimuli, corresponding to 12 targets, in each trial. Within one block, each trial marks a different visual stimulus. %In summary, by the end of the experiment, each subject would have looked at 12 different targets, each target for 4 seconds, and a total of 15 times per target. 
% The recorded raw EEG data is down-sampled from 2048Hz to 256 Hz and band-pass filtered from 6 Hz to 80 Hz. The first 135 ms of samples of each trial is discarded, considering the latency delay in the visual system.

\subsubsection{Pre-processing and Feature Extraction}
\begin{figure}
    \centering
    \includegraphics[width=0.75\linewidth]{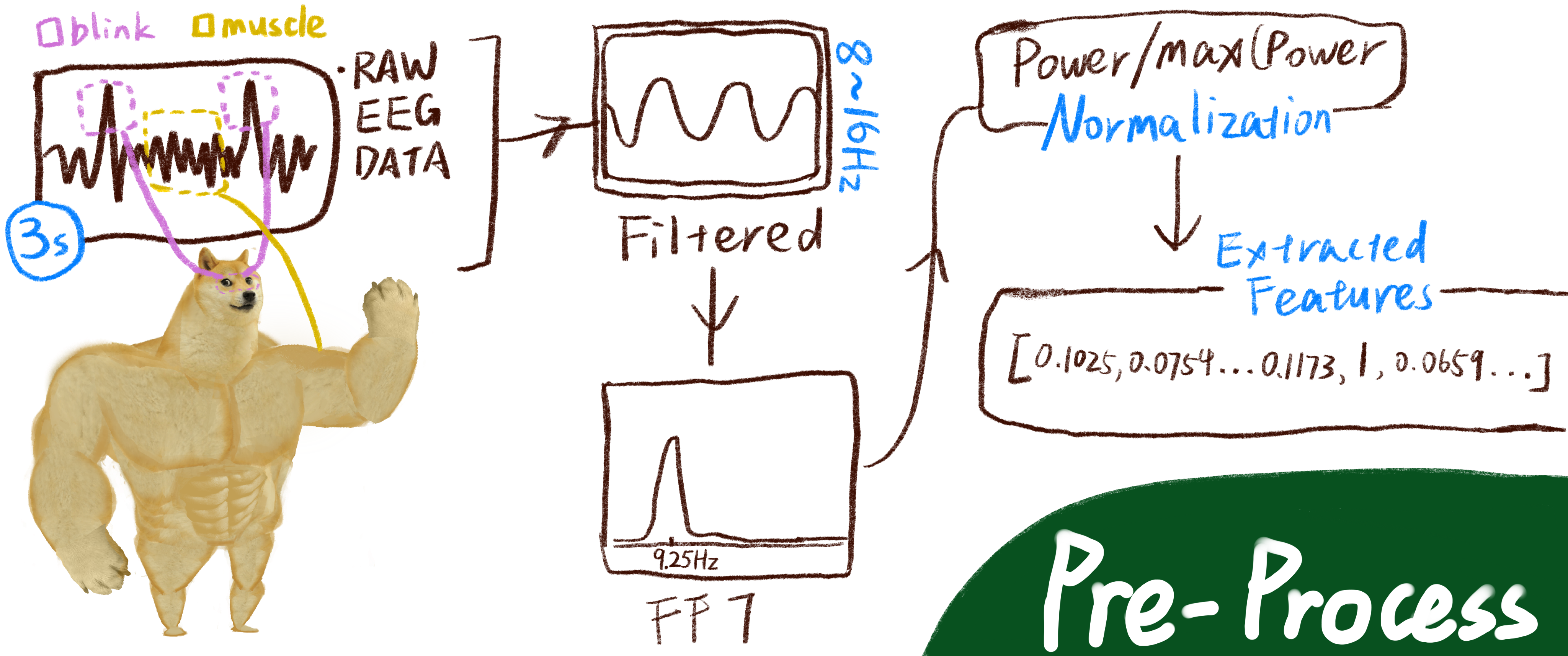}
    \caption{Pre-processing and feature extraction}
    \label{fig:pre}
\end{figure}

Due to the limitation of our EEG recording devices in online implementation, we only chose the channel "OZ" for doing classification. In the pre-processing stage, the EEG data is first filtered using a band-pass filter of 8 to 16 Hz, %6 order Butterworth band-pass filter with a lower cutoff frequency of 8 Hz and a higher cutoff frequency of 16 Hz,
encompassing our frequencies of interest: 9.25 Hz, 11.25 Hz, and 13.25 Hz. 
%This frequency cutoff range is usually considered optimal for SSVEP classification (Nakanishi et al., 2015). 
%Then a feature extraction process is implemented on the filtered data. 
To produce more data for training, the data is segmented into 3s overlapping windows with 16 samples offset. 
%Each window is 3 seconds long, containing 256*3 data points. The offset between each window is 16 samples. 
Fast Fourier Transform (FFT) is then applied to the filtered data in each window, to extract frequency features. The features are then normalized between 0 and 1 (Figure \ref{fig:pre}).
%transforming the data from time domain to frequency domain, enabling the analysis of frequencies. 
%Next, the magnitude corresponding to the frequency from 8 Hz to 16 Hz is then normalized between 0 and 1 (Figure \ref{fig:pre}).
% We choose the normalized power as features because it has been proven to be successful in doing SSVEP classification in Nguyen and Chung's previous work \cite{nguyen2018single}. 
%The magnitude corresponding to the frequency from 8 Hz to 16 Hz is extracted from the transformed data and normalized between 0 and 1. 
After extracting features from the three trials of interest (9.25Hz, 11.25Hz, 13.25Hz) of all blocks, there are 7650 labeled data points to be used for training. 80\% of the data is reserved for training and 20\% of the data is reserved for testing.

% How did you split the dataset, how much data is in train, validation and test. 

% why do you do this?
% last talk about the features we used

\subsubsection{Model Architecture and Training}

% We chose Convolutional Neural Network as the model due to its ability to ....
Our model consists two convolutional layers with 8 filters, each with size 3, a dropout layer with 25 \% dropout rate, a max-pooling layer with size 2, a flatten layer, and two dense layers with size 64 and 3. We adopted this structure from Nguyen and Chung's work and made slight adjustment to the hyperparameters based on empirical evidence. 
%(Fig. \ref{fig:CNN}).
% In the final model we deployed, we used 8 kernels with size 3 for both of the convolutional layers, each following a ReLu activation function. Next, we added a dropout layer with 25\% to prevent overfitting. Then we applied a maxpooling layer to reduce computation time. The first dense layer consists of 64 neurons, with the ReLu activation function. The second dense layer consists of 3 neurons, with soft max activation function, giving the output.  

%based on empirical evidence to better fit our purpose, computational resources, and dataset. For example, we changed the number of filters from 10 to 8, and kernel size from 5 to 3. So each of our two convolutional layers has 8 3-point kernel filters. ReLu is used as the activation function. A dropout layer with the rate of 25\% is added to prevent the model from overfitting. Then we applied a maxpooling layer to reduce computation time. The first dense layer consists of 64 neurons, with the ReLu activation function. The second dense layer consists of 3 neurons, with soft max activation function, giving the output. 

% During the training process, since our data is limited, we applied 5-fold cross validation to better evaluate the model performance. We also plot the loss and accuracy for training and validation to monitor and prevent overfitting.

The training is completed on a MacBook Air laptop with a 1.1GHz quad-core Intel Core i5 processor and 8GB of 3733MHz LPDDR4X onboard memory. The computation time to train the model is roughly 60 seconds.

% give reasons to why we chose those hyperparameter, see the paper I share "one channel..."

% Specify computational resources: type of computer, memory, cpu etc. also specify how long for the training

% talk about the 5-fold cross-validation thing we did. Why did we do it? What is the result? Is the validation accuracy what we expected them to be? Are we overfitting, underfitting etc.?

% \begin{figure}
%     \centering
%     \includegraphics[width=1\linewidth]{Styles/model.png}
%     \caption{CNN model architecture}
%     \label{fig:CNN}
% \end{figure}

\subsubsection{Robot autonomous navigation algorithm and wireless connection module}
\begin{figure}
    \centering
    \includegraphics[width=0.75\linewidth]{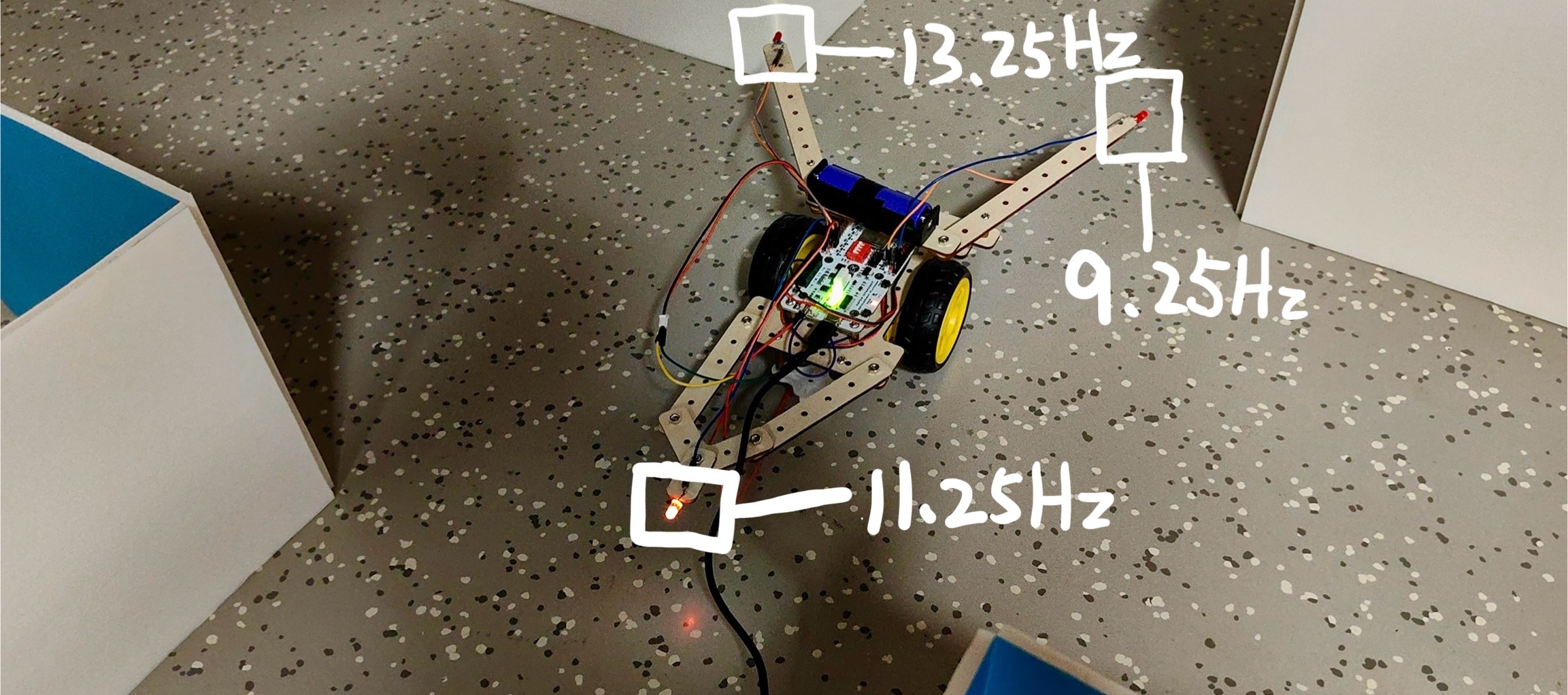}
    \caption{Robot car halting at a cross-section, waiting for command}
    \label{fig:car}
\end{figure}

% Describe the autonomous navigation algorithm, it's basically what the robot does when there is no selection to make. Preferrably draw a state diagram
Due to time constraints, we only implemented the wireless connection module, but have not implemented the navigation module. But here is how we would do it given sufficient time: the car will be equipped with 3 ultrasonic sensors in the front and both side that constantly scans the distance to determine if the car is able to turn in any direction. If the car has more than one way to go, it will first halt, then trigger the 3 LEDs to flash at the predetermined frequency. After waiting for 3 seconds, the car will get the result from the Python server in which the SSVEP classification results are stored. After data retrieval, the car carries out the corresponding command, moves forward, and continues to determine if turning is possible.
Figure \ref{fig:car} shows the car halting at a cross-section and waiting for commands.

%Figure. x shows the state diagram of the navigation algorithm. 
%when the car enter to a maze, the car will start to determine whether it is impassable or can turn immediately. If the car is facing a situation which is able to turn, the car will get dataset from the server and get the SSVEP data to decide whether the car will go straight om or turn whether right or left. The python server can get the data from the computer and the OpenBCI cyton board to get the SSVEP data to give command to the car.

% \begin{figure}
%     \centering
%     \includegraphics[width=0.75\linewidth]{企业微信截图_20240626214914.png}
%     \caption{???}
%     \label{fig:enter-label}
% \end{figure}

\subsection{Online implementation}
In the online implementation phase, we tested the feasibility and effectiveness of our proposed system. The results proved that our prototype system to be fully functional, that is, we are able to deploy the pre-trained model, get classification result every three seconds, and send the classification result to the robot through wireless connection. The robot, carried out the correct command based on the received data 0,1,2, and after the command is carried out, the robot turned on the LEDs and waited for another command. The video demonstration of our system running is in the Github in the appendix section. 
%Figure. X shows a graphical representation of the online implementation. 

% The text must be confined within a rectangle 5.5~inches (33~picas) wide and
% 9~inches (54~picas) long. The left margin is 1.5~inch (9~picas).  Use 10~point
% type with a vertical spacing (leading) of 11~points.  Times New Roman is the
% preferred typeface throughout, and will be selected for you by default.
% Paragraphs are separated by \nicefrac{1}{2}~line space (5.5 points), with no
% indentation.

% The paper title should be 17~point, initial caps/lower case, bold, centered
% between two horizontal rules. The top rule should be 4~points thick and the
% bottom rule should be 1~point thick. Allow \nicefrac{1}{4}~inch space above and
% below the title to rules. All pages should start at 1~inch (6~picas) from the
% top of the page.

% For the final version, authors' names are set in boldface, and each name is
% centered above the corresponding address. The lead author's name is to be listed
% first (left-most), and the co-authors' names (if different address) are set to
% follow. If there is only one co-author, list both author and co-author side by
% side.

% Please pay special attention to the instructions in Section \ref{others}
% regarding figures, tables, acknowledgments, and references.

\section{Result}
\subsection{Pre-trained model result}

\begin{figure}
    \centering
    \includegraphics[width=0.75\linewidth]{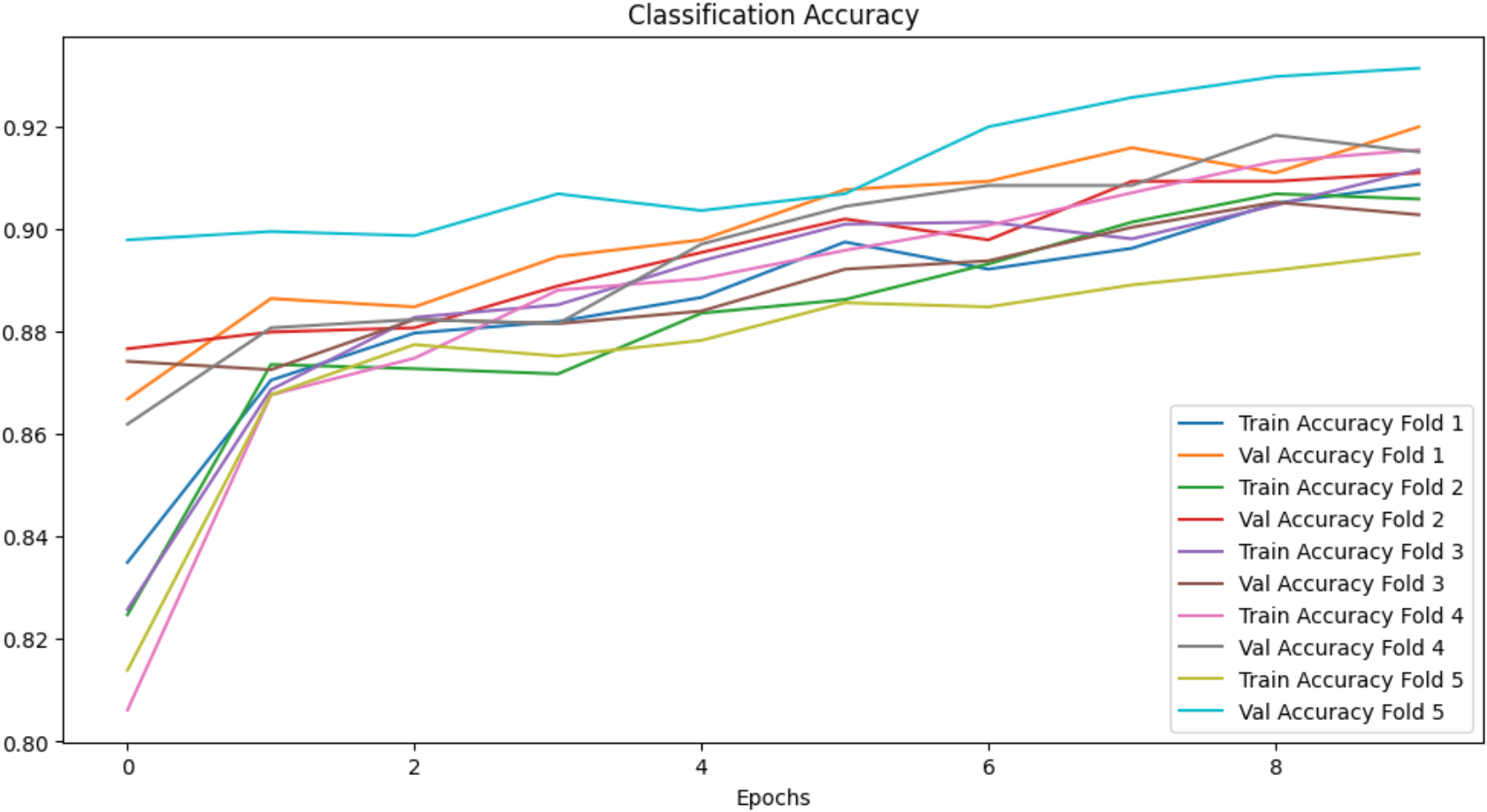}
    \caption{5-fold cross validation results for training and validation with 3s window length}
    \label{fig:acc}
\end{figure}

Figure \ref{fig:acc} displays the 5-fold cross-validation accuracy for training and validation with 3s window length. The test set accuracy for that model is 94.575\%. The model with the data window length of 2 seconds gives a test accuracy of 84.983\%. The test accuracy for the model with a 1-second window length is 76.168\%. 

%Plot 1s, 2s, 3s classification accuracy
%How did our robot perform in the task? What are some data we can use to quantify how good our system is? 

\subsection{Online Model deployment result}
We deployed the pre-trained model online and tested the system with our EEG recording devices. Despite the system proved to be fully functional, that is the car being able to receive the command through wireless connection and carry out appropriate actions, our online classification accuracy is much lower than expected. The "move forward" command reached a accuracy of 80\%, while the other two commands is roughly at chance level. There are several reasons we suspect that cause such a discrepancy between offline training and online testing result. First, our EEG recording electrode and device is extremely cheap, roughly \$150 combined, which means it could have a much worse signal to noise ratio as compared to the one other researchers use. Second, we did not train the model using our dataset, but instead using a different dataset that was recorded on a different device. We did not expect the results to differ than much because we thought the features for SSVEP should be applicable across devices. However, The results did not support our hypothesis. Other reasons might be because the car is a moving object, so when the subject shifts his/her gaze to look at another LED, there is inevitably some muscle movement happening, decreasing the classification accuracy. 
\section{Discussion}
Our inspiration to build this neural-controlled robot rises from Zhang's work \citep{zhang2023noir}. In their work, they built a much more comprehensive robot controlled by neural signals that is capable of carrying out 20 daily activities. Compared to our robot, their robot is definitely more general-purpose and can carry out much more tasks. However, the cost of building our robot is approximately 110 Yuan, or 15 dollars, which includes the ESP32 microcontroller, the LEDs, the ultrasound modules, and the battery pack. The low cost of our design allows hobbyists and designers to quickly replicate our work and make improvements to it. We believe that as SSVEP classification algorithms continue to improve, dry EEG electrodes become cheaper and have better signal-to-noise ratio \citep{liu2023feature}, and robot path-finding algorithms become smarter, the idea of implementing our design at elderly homes for daily activity assistance will one day become a reality. 

\begin{ack}
% Use unnumbered first-level headings for the acknowledgments. All acknowledgments
% go at the end of the paper before the list of references. Moreover, you are required to declare
% funding (financial activities supporting the submitted work) and competing interests (related financial activities outside the submitted work).
% More information about this disclosure can be found at: \url{https://neurips.cc/Conferences/2024/PaperInformation/FundingDisclosure}.

The authors thank Dr. Masaki Nakanishi for his kind assistance on helping us understand his dataset. The authors would also to like to thank Dr. Tzyy-Ping Jung for providing his valuable insight on doing SSVEP classification. The authors thank Dr. Wei Du, Yang Guo, Vernon Peens, Fangyu Liu, Zherong Gu for kindly supporting us with the equipment to accomplish this project. 
\end{ack}

%\section*{References}

% References follow the acknowledgments. Use unnumbered first-level heading for
% the references. Any choice of citation style is acceptable as long as you are
% consistent. It is permissible to reduce the font size to \verb+small+ (9 point)
% when listing the references.
% Note that the Reference section does not count towards the page limit.
\bibliographystyle{plainnat}
\bibliography{neurips_hs_2024.bib}
% [1] Alexander, J.A.\ \& Mozer, M.C.\ (1995) Template-based algorithms for
% connectionist rule extraction. In G.\ Tesauro, D.S.\ Touretzky and T.K.\ Leen
% (eds.), {\it Advances in Neural Information Processing Systems 7},
% pp.\ 609--616. Cambridge, MA: MIT Press.

% [2] Zhang, Ruohan, et al. "Noir: Neural signal operated intelligent robots for everyday activities." arXiv preprint arXiv:2311.01454 (2023).

% [2] Bower, J.M.\ \& Beeman, D.\ (1995) {\it The Book of GENESIS: Exploring
%   Realistic Neural Models with the GEneral NEural SImulation System.}  New York:
% TELOS/Springer--Verlag.

% [3] Hasselmo, M.E., Schnell, E.\ \& Barkai, E.\ (1995) Dynamics of learning and
% recall at excitatory recurrent synapses and cholinergic modulation in rat
% hippocampal region CA3. {\it Journal of Neuroscience} {\bf 15}(7):5249-5262.
% }
\appendix
\section{Appendix}
All materials needed to reproduce the experimental results, including code to train the model, the Arduino code to control the robot car, are uploaded to \url{https://github.com/QABCI/Neural-Signal-Operated-Intelligent-Robot}
\end{document}